\documentstyle[preprint,prl,aps]{revtex}
\tightenlines
\long\def\title#1{{\Large\begin{center}#1\end{center}\par}}

\long\def\address#1{\begin{center}#1\end{center}\par}

\long\def\author#1{\begin{center}#1\end{center}\par}

\def\pacs{}

\begin{document}

\draft

\title {Stripe phase ordering as quantum interference phenomenon.}

\author{S.I. Mukhin} 
\address{\cite{M}Theoretical Physics Dept.,
Moscow Institute for Steel and Alloys, Leninskii pr. 4,
117936 Moscow, Russia\\
Lorentz Institute for Theoretical Physics, Leiden University, 2300 RA Leiden,
The Netherlands}

\date{\today}
\maketitle

\begin{abstract}

A quantum interference mechanism of the stripe phase instability in quasi
one-dimensional (1D) repulsive electron system is proposed. The leading
spin-charge coupling term in Landau functional is derived microscopically.
It is shown that away from half filling, periodic lattice potential causes
cooperative condensation of the spin and charge superlattices, which
constitute a 1D analogue of the experimentally observed quasi 2D structures
in lanthanum cuprates and nickelates. Renormalization group analysis 
qualitatively
supports the Hartree-Fock picture. 
\end {abstract}

\pacs{PACS numbers: 64.60.-i, 71.27.+a, 74.72.-h, 75.10.-b}

Stripe phases recently observed in doped antiferromagnets \cite{1} attract attention 
to
the problem of multi-mode instabilities in the interacting electron systems
\cite{2}-\cite{4}.
Spin and charge modulations found in the doped layered transition metal oxides possess
locked mutual phase and doping dependent periods, incommensurate with the period of the
underlying crystal lattice. Period of the spin modulation is twice the period of the
charge modulation. An onset of this picture in CuO$_{2}$ and NiO$_{2}$ atomic planes is
detected by neutron scattering experiments \cite{1} which, below some temperature,
reveal two systems of intense superlattice peaks. Locations of the spin and charge 
peaks in the two-dimensional square Brillouin zone are: $(1/2\pm\epsilon ,1/2)$ for
spin and $(\pm 2\epsilon ,0)$ for charge; as well as at $\pi/2$ rotated positions:
$(1/2,1/2\pm\epsilon)$ and $(0,\pm 2\epsilon)$. Here units of $2\pi/a$ are used, and
$a$ is the lattice period. For small enough doping concentration, $x_{d}$,
a relation holds: $\epsilon\sim x_{d}$, and at greater $x_{d}$'s a shoulder developes.
Commensurate antiferromagnetic order at half filling (i.e. at $x_{d}=0$) is
characterized with the wave vector $\vec{Q}_{AF}=(1/2,1/2)$. Existing theoretical 
works on the stripe phase instability either find variationally the ground state of
the microscopic models \cite{2,3}, or use phenomenological Landau-Ginzburg functional
constructed by symmetry considerations \cite{4}.

In this Letter we propose a quantum interference mechanism of the stripe phase
instability in (quasi) one-dimensional (1D) electron system, brought about by the
periodic lattice potential. A microscopic derivation is presented of the main 
spin-charge coupling term in the Landau functional, which was phenomenologically
introduced in \cite{4}. A model considered below was already investigated in the
Hartree-Fock approximation \cite{3}, but proposed here mechanism of the stripe phase
ordering was not discussed. Neither was it discussed in the previous studies of the
1D electron systems within one-loop renormalization group approach \cite{5}
("parquet" formalism), which is considered in the last part of this Letter.

We start from the Hartree-Fock approximation. The Hubbard Hamiltonian with the hopping
integral $t$ and on-site repulsion $U$ ($>0$) may be written in the form:

\begin{equation}
H=\displaystyle t\sum_{\langle i,j\rangle \sigma}c^{\dagger}_{i,\sigma}c_{j,\sigma}+
U\displaystyle\sum_{i}\left(\frac{1}{4}\hat{n}^{2}_{i}-(\hat{S}^{z}_{i})^{2}\right)
\label{hubbard}
\end{equation}

\noindent
Here an identity:
$\hat{n}_{i\uparrow}\hat{n}_{i\downarrow}=\frac{1}{4}\hat{n}^{2}_{i}-
(\hat{S}^{z}_{i})^{2}$ was used. Operators
$\hat{n}\equiv\hat{n}_{\uparrow}+\hat{n}_{\downarrow}$ and $\hat{S}^{z}$ are fermionic
density and spin (z-component) operators respectively, and $\sigma$ is spin index. 
Hamiltonian (\ref{hubbard}) 
has convenient form for the Hartree-Fock decoupling in the presence of two, i.e. spin
and charge, order parameters. At half filling, i.e. at one electron per site, the 
system orders antiferromagnetically \cite{6,7}:
$S^{z}_{i}=(m_{o}/2)\cos{(\vec{Q}_{AF}\vec{r}_{i})}$. Away from half filling, 
mean-field
approximation \cite{8} predicts a spin density wave (SDW) instability in the
continuous quasi 1D system: $S^{z}(x)=(m_{o}/2)
\cos{(Q_{-}x-\phi)}$, with the incommensurate wave vector $Q_{-}=2k_{F}$. In the 
weak coupling limit, $U\ll t$, the SDW instability is brought about by the existence
of the nested parts of the Fermi surface separated by momentum $2k_{F}$. The
difference $Q_{AF}-Q_{-}\equiv 1/2-2k_{F}=\epsilon$ is expressed via the hole 
(electron) doping concentration $x_{d}>0$ ($x_{d}<0$) as: $\epsilon = x_{d}/2$.
Effect of the periodic lattice potential away from half filling has been so far
neglected, both in the mean-field and in the "parquet" approaches \cite{8,5}.

Consider an incommensurate SDW expression, $S^{z}(x_{i})=(m_{o}/2)
\cos{(Q_{-}x_{i}-\phi)}$, on a chain. Here $x_{i}$ is the "site coordinate"
along the chain, and $\phi$ is a phase shift. Allowing for the periodicity of the 
lattice
sites, $x_{i}=N_{i}a$; $N_{i}=0,\pm 1,\pm 2, ...$, we may rewrite the cosine in 
the SDW
expression in the equivalent form ($x_{i}$ is taken in units of $a$):
$\cos{(Q_{-}x_{i}-\phi)}= (1/2)(\cos{(Q_{-}x_{i}-\phi)}+\cos{(Q_{+}x_{i}+\phi)})
=\cos{(x_{i}/2)}\cos{(\epsilon x_{i}+\phi)}$; where $Q_{+}=2\pi/a-Q_{-}$, i.e.
$1/2+\epsilon$ in our dimensionless units. These simple relations demonstrate 
significant physical fact that an incommensurate SDW on the lattice can be decomposed
into two umklapp related SDW's, which could be called "direct" $(Q_{-}=2k_{F})$ and
"shadow" $(Q_{+}=2\pi/a-2k_{F})$ waves \cite{9}. It is wellknown \cite{10} that
formation of SDW with wave vector $Q$ may be regarded as Bose-condensation of 
electron-hole pairs $c_{k,\sigma}c^{\dagger}_{k+Q,\sigma}|O\rangle$ with momentum $Q$
($|O\rangle$ is unperturbed vacuum state of the Fermi-system). In our case this means
formation of the {\it{two}} electron-hole condensates, with wave vectors $Q_{\pm}$,
corresponding to "direct" and "shadow" SDW's. Then, scattering of electrons (holes)
by some "extra" periodic potential with "matching" wave vector
 $Q_{+}-Q_{-}\equiv 2\epsilon$ would cause quantum interference between the wave 
functions of the $Q_{\pm}$ condensates. At some value of the relative phase shift the
interference may become constructive, causing enhancement of the gap in the fermionic
quasi-particle spectrum of the quasi 1D electrons and related decrease of their total
(free) energy. Here we consider a charge density wave (CDW): $\rho(x)=\rho_{o}
\cos{(Q_{+}-Q_{-})x}$, as an option for the "matching" extra potential. Depending
on the balance between the gain of energy and the "cost" of the CDW formation, a
condensation of the coupled SDW-CDW phase takes place as a first/second order phase
transition. This state, consisting of "direct" and "shadow" SDW's and "matching" CDW,
with locked relative phase shift $\phi$, is right a weak coupling analogue of the 
stripe phase order considered in the literature \cite{1}-\cite{4}.

In the presence of the SDW and CDW condensates, i.e. densities $m(x)$ and $\rho(x)$
introduced above, the single-particle eigenstates of the Hamiltonian Eq.(\ref{hubbard})
in the Hartree-Fock approximation can be determined from the Bogoliubov-de Gennes
equations derived in \cite{3}:

\begin{eqnarray}
\mp i2t\frac{\partial u_{\pm}}{\partial x} + 
\frac{U}{2}\rho(x)u_{\pm}-\frac{U}{2}\tilde{m}(x)u_{\mp}= Eu_{\pm}
\label{Gennes}
\end{eqnarray}

\noindent
Here left- and right-movers representation is used for the quasi-particle wave
function: $\psi_{\sigma}(x)=u_{+}(x)\exp{(ix/4)}+\sigma u_{-}(x)\exp{(-ix/4)}$,
 and coordinate 
$x$ is expressed in units of the chain period, $a$. The spin density is:
$m(x_{i})/2=(m_{o}/2)\cos{(Q_{-}x_{i}-\phi)}=\tilde{m}(x_{i})\cos{(x_{i}/2)}$, so 
that: $\tilde{m}(x)
\equiv(m_{o}/2)\cos{(\epsilon x+\phi)}$. Thus, only slowly varying functions
 $u_{\pm}(x)$, $\tilde{m}(x)$ and $\rho(x)$ are involved in Eq.(\ref{Gennes}). Now,
instead of dropping CDW term, $\rho(x)u_{\pm}$, from Eq.(\ref{Gennes}) \cite{3}, we
shall explicitly allow for it by writing wave functions $u_{\pm}(x)$ in the Bloch
wave basis of the periodic CDW potential:

\noindent
$u_{\pm}(x)\equiv c^{k}_{\pm}\exp{(ikx\mp iU/(2v)\int^{x}\rho(x')dx')}
\approx c^{k}_{\pm}e^{ikx}\left\{J_{0}(z)\mp J_{1}(z)
\left(e^{i2\epsilon x}-e^{-i2\epsilon x}\right)\right\}$;  

\noindent
where $v=2t$ is the Fermi velocity of electrons, and $z=U\rho_{o}/(2v\epsilon)$.
Here $J_{n}$ is the Bessel function of the integer order $n$, and the terms of
higher orders than $n=1$ are neglected, provided that $U\rho_{o}/(2v\epsilon)\leq 1$.
After substitution of $u_{\pm}(x)$ in the above representation into 
Eq. (\ref{Gennes}) one finds an algebraic system of linear homogeneous equations for
the coefficients $c^{k}_{\pm}$. Solving corresponding determinant equation one finds
the single-particle spectrum:

\begin{eqnarray}
&&E_{k} = \displaystyle -\frac{v\epsilon}{2}\pm
\sqrt{\left(k+\frac{\epsilon}{2}\right)^{2}v^{2}+\Delta^{2}};
                                                        \label{spec}
\\
&&\Delta\equiv\frac{Um_{o}}{4}f\left(\frac{U\rho_{o}}{2v\epsilon}\right)\nonumber\\
&&f(z)^{2}\equiv J_{0}^{2}(z)-2\cos(2\phi)J_{0}(z)J_{1}(z)+J_{1}^{2}(z)
                                                        \label{f}
\end{eqnarray}

\noindent
In the "electron doping" case the sign in front of $\epsilon$ in Eq.(\ref{spec})
and of $\cos{2\phi}$ in Eq. (\ref{f}) should be changed. 
The physical implication of Eqs. (\ref{spec}), (\ref{f}) is remarkable. Namely,
in the presence of the CDW the effective coupling constant $Uf(z)$, responsible for
the SDW condensation, is enhanced with respect to bare coupling $U$, provided $f(z)>1$.
The latter condition fixes $\cos(2\phi)$. The form of Eq.(\ref{f}) manifests 
quantum interference between scattering amplitudes of electron in the combined
periodic potentials of $Q_{\pm}$-SDW and $(Q_{+}-Q_{-})$ CDW. A simple form of 
solution
(\ref{spec}) is valid in the weak coupling limit, $U\ll t$, not too close to half 
filling, i.e. when $x_{d}\gg \Delta/t$. Also, despite the gap in the spectrum
(\ref{spec}), the system may remain conductive due to sliding of the incommensurate 
density waves along the lattice.

Free energy of the system (per unit of length), $\Omega$, at finite temperature
$T(\equiv\beta^{-1})$, follows from Eq.(\ref{spec}) and the Hartree-Fock form 
\cite{3}
of the Hamiltonian (\ref{hubbard}):

\begin{eqnarray}
\Omega = \displaystyle (U/8)(m_{o}^{2}/2 +\rho_{o}^{2})-(4T/\pi v)
\displaystyle\int^{E_{b}}_{0}
\ln{\left[2\cosh{(\beta E(\xi)/2)}\right]}\, d\xi
\label{energy}
\end{eqnarray}

\noindent
where $E(\xi)=\sqrt{\xi^{2}+\Delta^{2}}$, and $E_{b}(\sim t)$ is the upper cutoff of
the electron energy. Due to interference term $-2\cos(2\phi)J_{1}(U\rho_{o}/
2v\epsilon)J_{0}$ in Eq.(\ref{f}), the lowest order expansion of $\Omega$, 
Eq.(\ref{energy}), in powers
of the CDW amplitude (at small $m_{o}$) yields linear in $\rho_{o}$ term:
 $\delta\Omega\sim -m^{2}_{o}\rho_{o}U^{2}/(tx_{d})$. This term reveals possible
microscopic origin of the corresponding spin-charge coupling term in the 
phenomenological Landau-Ginzburg functional considered in \cite{4}.

Here we merely list the main results following from Eq.(\ref{energy}).

i) Coming from the high temperature limit, $\Delta =0$, the stripe phase
condenses first with
$\cos{(2\phi)}=-1$ or $\cos{(2\phi)}=1$ depending on the sign of $x_{d}$. 
Thus, the spin density behaves as:
 $\langle2S^{z}(x)\rangle=
-m_{o}\cos{(x/2)}\sin{(\epsilon x)}$ in the case $x_{d}>0$, and as : 
$\langle2S^{z}(x)\rangle=
m_{o}\cos{(x/2)}\cos{(\epsilon x)}$ in the $x_{d}<0$ case. Simultaneously, electron 
charge density is the same in both cases: $-\rho_{o}\cos{(2\epsilon x)}$.
Hence, the nodes of the spin density coincide with the minima (maxima) of the charge
 density $\rho(x)$ in the case of the hole (electron) doping. This topology is in 
accord with the stripe phase topology considered in the strong coupling limit 
\cite{2}-\cite{4}.

ii) When doping concentration $x_{d}$ decreases below $ x_{o}\sim\sqrt{t/U}
\exp{(-2\pi t/U)}$ the system enters strong coupling regime, see Figs. 1,2. This
transition is governed by dimensionless parameter :$U\rho_{o}/(2\pi tx_{d})$.
At $U\rho_{o}/(2\pi tx_{d})\geq 1$ higher order Bessel functions contribute to the
wave function $u_{\pm}(x)$ and a few-harmonic approximation, resulted in
Eqs.(\ref{spec}), (\ref{energy}), fails. This indicates a solitonic, rather than
SDW-CDW structure of the stripe phase in the lowest doping limit, 
$x_{d}\rightarrow 0$,
in a qualitative accord with the previous proposals \cite{2,3}. Stripe phase
condensation temperature, T$_{c}$, monotonically decreases from the {\it{highest}}
value $T_{m}= 2(\gamma/\pi)t\exp{(-2\pi t/(Uf^{2}_{m}))}$ at small doping 
concentrations, $|x_{d}|<x_{o}$, to the {\it{lowest}} value 
$T_{SDW}= 2(\gamma/\pi)t\exp{(-2\pi t/U)}$ at $|x_{d}|\gg x_{o}$. Here $\gamma=1.78$,
and $f_{m}\equiv f(z_{m})\approx 1.2$ is the maximum value of the function $f(z)$ in
Eq.(\ref{f}), reached at $z=z_{m}\approx 0.83$. The drop of T$_{c}$ is accompanied by
a substantial decrease of the SDW and CDW amplitudes at zero temperature, see Fig. 1: 

\noindent
$m_{o}= 8t/U \exp{(-2\pi t/U)}$, at $x_{d}\gg x_{o}$; and 
$m_{o}=8t/(Uf_{m})\exp{(-2\pi t/(Uf_{m}^{2}))}$, at $x_{d}\ll x_{o}$.

\noindent
Simultaneously, the character of the phase transition changes at $x_{o}$ from the 
{\it{first order}} ($x_{d}<x_{o}$) to the {\it{second order}}, Fig. 2. The jumps of 
the CDW and SDW amplitudes at the first order transition temperature are: 
$m_{o}^{2}\approx x_{d}z_{m}T_{m}\sqrt{2\pi tU}/(Uf_{m})^{2}$ and 
$\rho_{o}\approx 2\pi x_{d}z_{m}t/U$. The latter estimate for $\rho_{o}$ results in
the following value of the "small parameter": $U\rho_{o}/(2\pi tx_{d})\approx 0.83
<1$. Hence, our results in the region $x_{d}<x_{o}$, based on the neglect of the 
higher order SDW/CDW harmonics, might be considered as qualitative rather than
quantitative.

iii) In the second order phase transition regime, $|x_{d}|\gg x_{o}$, the order 
parameters close to $T_{c}$ behave as: $\rho_{o}\approx
3\tau T_{SDW}^{2}/(x_{d}tU)$, and $m_{o}\approx 12\sqrt{\tau}T_{SDW}/U$; in 
qualitative accord 
with \cite{4} (here $\tau\equiv 1-T/T_{SDW}$). In the classification of \cite{4} the
phase transitions described in ii), iii) above belong to the "spin-charge coupling 
driven" and "spin driven" kinds.

Important issue for the (quasi) 1D systems is the influence of fluctuations. We
study it within a single-loop renormalization group (RG) scheme, so-called
"parquet" approximation \cite{5}, which we adjust for the case of the two order
parameters (SDW/CDW) coupled already on the mean-field level. Conventionally,
"parquet" - RG equations describe behavior of the two-electron scattering vertices
$\gamma_{1}(\xi)$, $\gamma_{2}(\xi)$, and $\gamma_{3}(\xi)$, accounting for  back-, 
forward- , and umklapp scattering of electrons respectively close to the
Fermi "surface" points: $\pm k_{F}$. The RG variable, $\xi$, is the logarithm of the
infrared cutoff of the energy/momentum transfer. It is involved in the 
(logarithmically)
diverging corrections to the vertices, which are initially defined in the Born 
approximation: $g_{i}\equiv \gamma_{i}(\xi=0)$. Within "parquet" approach only
corrections of the highest power in $\xi$ are retained in each order of the 
perturbation
expansion in each $\gamma_{i}$ and then summed to an infinite order. An instability 
of
the electron system is manifested by divergences of the (initially finite) vertices
$\gamma_{i}(\xi)$ at some finite value $\xi_{o}$. In the case of the Hubbard 
Hamiltonian (\ref{hubbard}) at half filling: $g_{i}=U/(4\pi t)$, $i=1,2,3$,
in the dimensionless units. Away from half filling the umklapp condition:
$p_{1}+p_{2}=p_{3}+p_{4}\pm 2\pi/a$, can not be fulfilled when all the quasi-momenta
of electrons (before and after scattering) are close to the Fermi surface. In 
conventional "parquet" theory \cite{5} an instability does not occur in repulsive 1D 
system without umklapp, i.e. when $g_{1}, g_{2}>0$, $g_{3}=0$. In the (hole) doped 
case
the deficiency of momentum transfer: $2\pi/a-4k_{F}\equiv 2\epsilon=x_{d}\neq 0$,
provides a "natural" momentum cutoff, such that at $\xi>\xi_{d}\equiv\ln{(1/x_{d})}$
the growth of $|\gamma_{i}(\xi)|$ terminates. Hence, in the standard approach \cite{5},
an instability at $\xi_{o}$ would be only possible if $\xi_{o}<\xi_{d}$, i.e. in the
low doping region: $x_{d}<x_{1}\equiv exp{(-\xi_{o})}$. In this region commensurate 
SDW is
indistinguishable from incommensurate one within a logarithmic accuracy of "parquet" 
with respect to the infrared momentum cutoff. Using estimate \cite{5}: 
$\xi_{o}\sim 2\pi t/U$, we find: $x_{1}=\exp{(-2\pi t/U)}$, which is inside our
mean-field "strong coupling" region: $x_{d}<x_{1}<x_{o}$. In order to probe the system 
for a stripe phase instability in the region: $x_{d}>x_{1}$, where incommensurability 
is well resolved, we modify "parquet" treatment by adding ficticious "stripe phase"
vertices $\tilde{\gamma}_{i}(\xi)$ with infinitesimal "starting" values 
$\tilde{\gamma}_{i}(\xi_{d})$ at $\xi_{d}<\xi_{o}$. The vertices 
$\tilde{\gamma}_{1,2}(\xi)$ describe "umklapp" scattering with the wave vector 
$2\epsilon=x_{d}$, brought by the CDW component of the stripe phase; while 
$\tilde{\gamma}_{3}(\xi)$ is due to combined lattice-CDW umklapp: $2\pi/a-2\epsilon$.
Thus, "enriched" RG-"parquet" equations in the interval $\xi_{d}<\xi<\infty$ become: 

\begin{eqnarray}
&&\dot{\gamma}_{3}(\xi)= -2\tilde{\gamma}_{3}(\xi)\tilde{\gamma}_{4}(\xi);\;
\dot{\tilde{\gamma}}_{4}(\xi)=-4 Re(\gamma_{3}(\xi)\tilde{\gamma}_{3}^{*}(\xi))
\nonumber\\
&&\dot{\gamma}_{4}(\xi)=-2\tilde{\gamma}_{3}(\xi)\tilde{\gamma}_{3}^{*}(\xi);\;
\dot{\tilde{\gamma}}_{3}(\xi)=-2\tilde{\gamma}_{3}(\xi)\gamma_{4}(\xi)
                                           \label{parquet}
\end{eqnarray}

\noindent
where $\gamma_{4}(\xi)=\gamma_{1}(\xi)-2\gamma_{2}(\xi)$, and same relation is valid 
between $\tilde{\gamma}_{4}$ and $\tilde{\gamma}_{1,2}$ \cite{11}. 
Diverging solutions of
Eqs.(\ref{parquet}) for two concentrations $x_{i}>x_{1}$ are shown in Fig. 3.
Analytical solution in the interval $\xi>\xi_{d}$ is: $\gamma_{3}(\xi)=
B\cosh{(C\ln{|\tanh{D(\xi-\tilde{\xi}_{o})}|}+\phi_{o})}$; $\tilde{\gamma}_{4}(\xi)=
\pm\sqrt{2(\gamma_{3}^{2}-B^{2})}$; $\gamma_{4}(\xi)=(D/2)
\coth{2D(\xi-\tilde{\xi}_{o})}$; $\tilde{\gamma}_{3}(\xi)=
DC/(\sqrt{2}\sinh{2D(\xi-\tilde{\xi}_{o})})$, where $\tilde{\xi}_{o}\sim \xi_{o}+
0.5(\xi_{o}-\xi_{d})\ln{(2/|\tilde{\gamma}_{i}(\xi_{d})|)}$, and all the constants
are determined from the boundary conditions for $\gamma_{i}$ and $\tilde{\gamma}_{i}$
at $\xi=0$ and $\xi\approx \xi_{d}$ respectively.

The divergence of the probe vertices $\tilde{\gamma}_{i}(\xi)$ signals \cite{5} in 
favour of spontaneous incommensurate (stripe) ordering in the ground state of the
system away from half filling. Though, neither mean-field, nor "parquet" approximation
gives decisive answer about the long-range order and/or gap in the quasi-particle
spectrum of 1D system \cite{5}.

In conclusion, a quantum interference mechanism of the stripe phase ordering in
repulsive (quasi) 1D electron system is proposed. Preliminary 1D renormalization
group ("parquet") analysis does not contradict the mean-field construction.
Substantial enhancement of the stripe ordering temperature due to interference
between charge and spin order amplitudes is found. Analysis of analogous possibility 
for the superconducting order parameter is now in progress.

Instructive discussions with Jan Zaanen, and useful comments by A.A. Abrikosov,
L.P.Gor'kov and Wim van Saarloos are highly 
appreciated. The work was supported in part by NWO and FOM (Dutch Foundation for
 Fundamental Research) during the author's stay in Leiden.

\newpage

\section{Figures}

\noindent
Fig.1. SDW ($m_{o}$) and CDW ($\rho_{o}$) stripe phase order parameters, as 
functions of doping concentration $x_{d}$ at zero temperature, calculated using
Eq.(\ref{energy}). Inset: normalized stripe phase transition temperature T$_{c}$
as function of doping. 4t/U=3.2.

\vspace{5mm}

\noindent
Fig.2. Calculated temperature dependences of the stripe phase order parameters,
$m_{o}$ (curves labeled with "$m$") and $\rho_{o}$ (curves not labeled) for the 
different doping concentrations $x$. Each pair of lines of the same type (e.g. solid, 
dashed, etc.) show $m_{o}$ and $\rho_{o}$ for each particular value of doping 
concentration $x$. 4t/U=3.2 .

\vspace{5mm}

\noindent
Fig.3. Solid lines: numerical solutions of the Eqs.(\ref{parquet}) for the vertices
$\gamma_{3,4}$ (curves 3 and 4 respectively) and $\tilde{\gamma}_{3,4}$
(curves 3'  and 4' respectively), for values of $\xi_{d}=\xi_{1,2}<\xi_{o}$.
Dashed lines: same as above, but for $\xi_{d}>\xi_{o}$. Starting values: $g_{1,2,3}=
-g_{4}=0.1$; $\tilde{\gamma}_{3}(\xi_{i})=-\tilde{\gamma}_{4}(\xi_{i})=0.01$. 


\begin{thebibliography}{99}
\bibitem{9} We found that in a continuous periodic lattice potential (not site)
the "shadow" wave has finite weight, $\theta <1$, which vanishes together with Fourier
coefficient $\alpha_{\pm}$ in the electron (Bloch) wave function: $\psi_{k_{F}}(x)=
\exp{(ik_{F}x)}(1+\alpha_{\pm}\exp{(\pm i2\pi x/a)}+...)$.
\bibitem{11} We omit equations for the vertices $\gamma_{1}$ and $\tilde{\gamma}_{1}$, 
which do not diverge in the repulsive case considered here. A study of the effect of
introduction of the "superconducting" probe vertices, $\tilde{\gamma}$, is in progress
now.
\end{thebibliography}
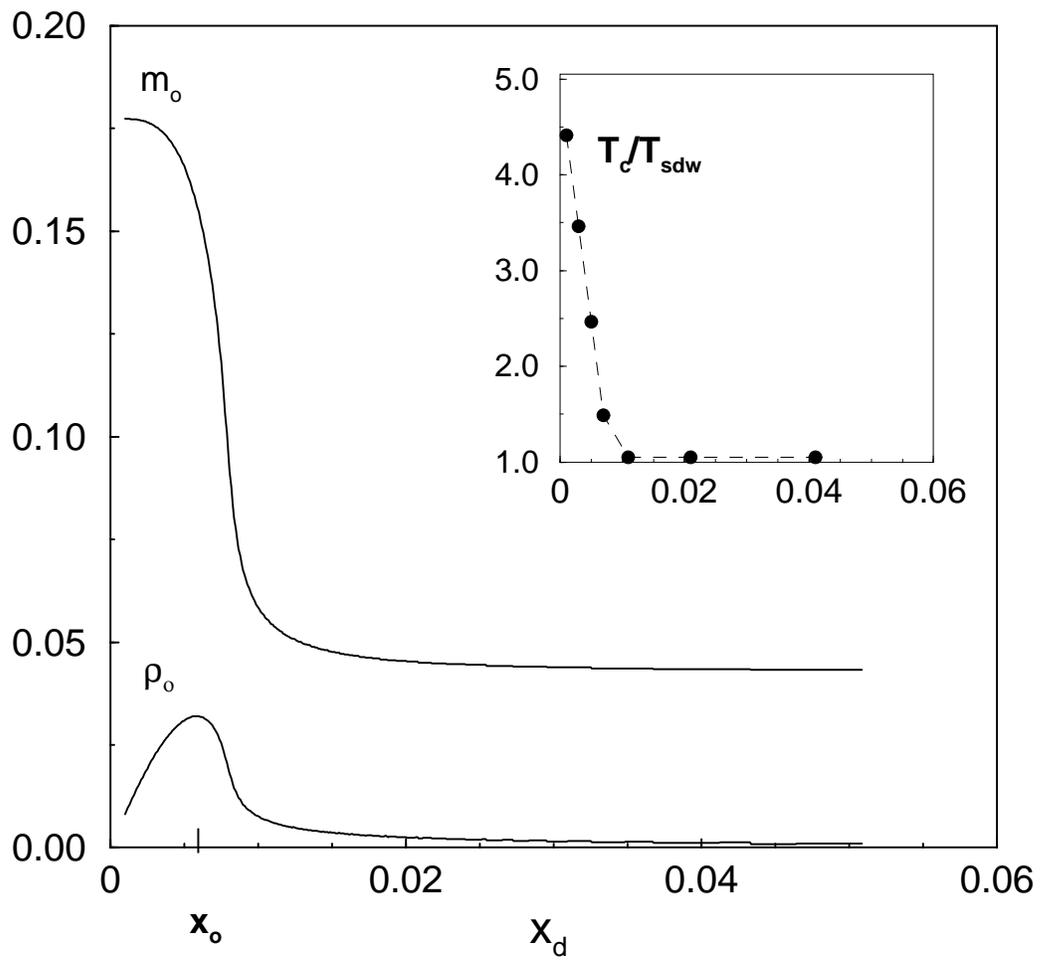
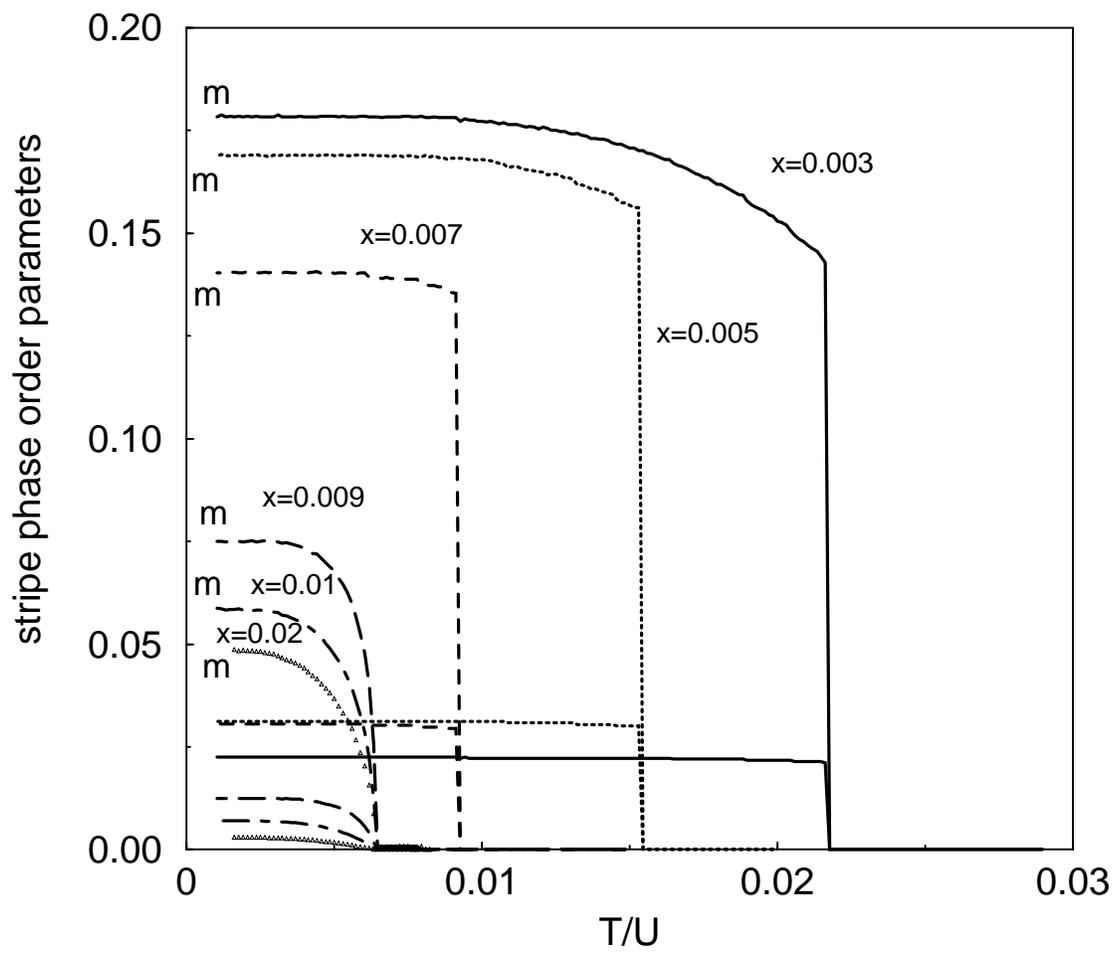
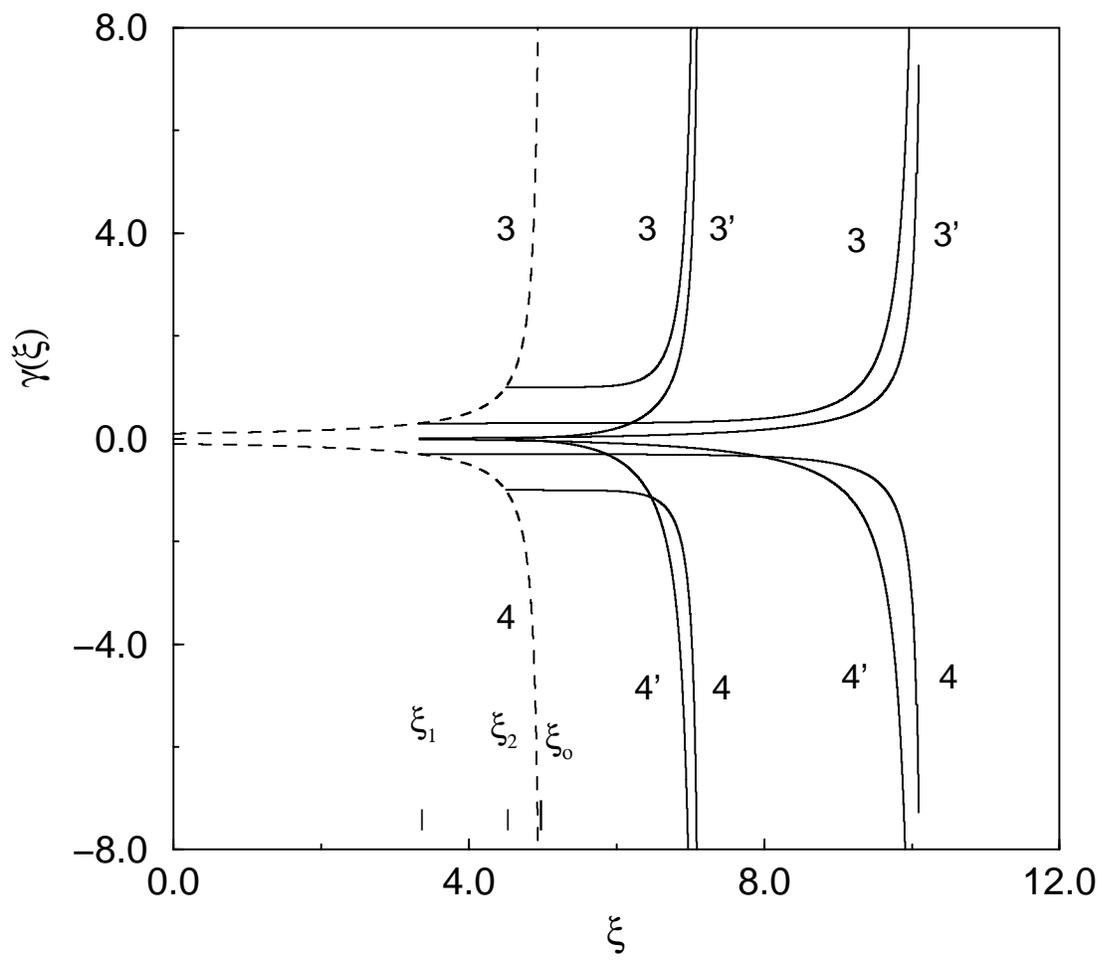
\end{document}